**REGULAR ARTICLE**  **Open Access**

# A large-scale community structure analysis in Facebook

Emilio Ferrara*

*Correspondence:
ferrarae@indiana.edu
Center for Complex Networks and
Systems Research, School of
Informatics and Computing, Indiana
University, Bloomington, USA
Department of Mathematics and
Informatics, University of Messina,
Messina, Italy

**Abstract**

Understanding social dynamics that govern human phenomena, such as communications and social relationships is a major problem in current *computational social sciences*. In particular, given the unprecedented success of *online social networks* (OSNs), in this paper we are concerned with the analysis of aggregation patterns and social dynamics occurring among users of the largest OSN as the date: Facebook. In detail, we discuss the mesoscopic features of the community structure of this network, considering the perspective of the communities, which has not yet been studied on such a large scale. To this purpose, we acquired a sample of this network containing millions of users and their social relationships; then, we unveiled the communities representing the aggregation units among which users gather and interact; finally, we analyzed the statistical features of such a network of communities, discovering and characterizing some specific organization patterns followed by individuals interacting in online social networks, that emerge considering different sampling techniques and clustering methodologies. This study provides some clues of the tendency of individuals to establish social interactions in online social networks that eventually contribute to building a well-connected social structure, and opens space for further social studies.

## Introduction

Social media and online social networks (OSNs) represent a revolution in Web users behavior that is spreading at an unprecedented rate during the latest years. Online users aggregate on platforms such as Facebook and Twitter creating large social networks of millions of persons that interact and group each other. People create social ties constituting groups based on existing relationships in real life, such as on relatives, friends, colleagues, or based on common interests, shared tastes, *etc.*

In the context of *computational social sciences*, the analysis of social dynamics, including the description of those unique features that characterize online social networks, is acquiring an increasing importance in current literature [1–3].

One of the challenges for *network scientists* is to provide techniques to collect [4] and process [5] data from online social networks in an automatic fashion, and strategies to unveil the features that characterize these types of complex networks [6]. In addition, these methods should be capable of working in such large-scale scenarios [7].

Amongst all the relevant problems in this area, the analysis of the so-called *community structure* of online social networks acquired relevant attention during latest years [8–14]. Recently, several relevant quantitative works have been presented to this purpose [15–19].

Studying the community structure of a network helps in explaining social dynamics of interaction among groups of individuals [20–22], but also to quantitatively investigate social theories such as Milgram's *small world* [23], Granovetter's *strength of weak ties* [24], Borgatti's and Everett's *core-periphery structure* [25, 26], and so forth.

Furthermore, discovering and analyzing the community structure is a topic of great interest for its economical and marketing implications [27]. For example, it could be possible to improve the advertising performance by identifying and targeting the most influential users of each community, exploiting effects such as the *word-of-mouth* and the spread of information within the community itself [28]. Similarly, exploiting the affiliations of users to communities might be effective to provide them useful recommendations on the base of common interests shared with their friends [29].

Finally, the community detection problem has plenty of challenges from a computational perspective, since it is highly related to the problem of clustering large, possibly heterogeneous, datasets [30–35].

In this work we are concerned with the analysis of the community structure of the largest online social network as to date: Facebook. In particular, we acquire a sample from the Facebook social graph (*i.e.,* the network of relationships among the users), and then we apply two different state-of-the-art algorithms to unveil its underlying community structure (see the Appendix for technical details).

The further analysis of the mesoscopic features of this network puts into evidence the organization patterns that describe the connectivity of users in large online social network.

To summarize, in the remainder of the paper we will discuss the following results:

(i) The emergence of a tendency of social network users at the formation of communities of heterogeneous size (following a heavy-tailed distribution), which means that there exist several groups of small size and a decreasing number of groups or larger size.

(ii) The number of interconnections that exists among communities also follows a broad distribution, that provides some clues in the direction of the assessment of the *strength of weak ties theory*, foreseen by the early work of Granovetter [24].

(iii) The community structure of the network is defined, independently of the method adopted to unveil it. To this purpose, we take into account the possible bias introduced by the sampling procedures [36] and the resolution limit suffered by some types of community detection algorithms [37, 38].

(iv) The emergence of the so-called *small world phenomenon* - whose existence in real-world social networks has been assessed during the sixties by Milgram [23]: the community structure of the network is highly clustered and tightly interconnected by means of short paths, features which are exhibited by several small world networks [15, 39]. According to the model of *small world network* proposed by Watts and Strogatz [39], not only the diameter of the network grows as the logarithm of the size (a feature exhibited also by random networks), but also the clustering coefficient is high - a discriminating feature observed also in our case.

## Methods

The aim of this work is to analyze the mesoscopic features of the community structure of the Facebook social network. In the following we provide some information about the process of data collection, briefly discussing the sampling methodology and the techniques adopted to collect data.

This is the first step to study the community structure of real-world networks, that reflect unique characteristics which are impossible to replicate by using synthetic network models [40].

After that, we discuss the process of community detection that we adopted to unveil the community structure of the network (and, to this regard, additional technical details are discussed in the Appendix).

Finally, we describe the process of definition of the community meta-network - a network whose nodes represent the communities identified in the social graph, to which it follows its analysis and discussion of findings.

**Sampling the Facebook network**

Differently from other online social network platforms (for example Twitter), Facebook does not provide a framework to automatically access information related to users with public profiles.

This lack of data availability has been faced acquiring public information directly from the platform, by means of a sampling process.

During this study we did not inspect, acquire or store personal information about users, since we were interested only in reconstructing the social connections among a sample of them - whose friend-lists were publicly accessible. To this purpose, we designed a Web data mining platform with the only ability to visit the publicly accessible friend-list Web pages of specific users, selected according to a sampling algorithm, and extract their connections. Obtained data have been used only to reconstruct the network sample studied in this work.

The architecture of the designed mining platform is briefly schematized as follows. We devised a data mining agent (*i.e.*, an autonomous software tool), which implements two sampling methodologies (*breadth-first search* and *uniform* sampling). The agent queries the Facebook server(s) in order to request the friend-list Web pages of specific users. In detail, the agent visits those Web pages containing the friend-list of a given user, following the directives of the chosen sampling methodology, and extracts the friendship relationships reported in the publicly accessible user profile.

The sampling procedure runs until any termination criterion/a is/are met (*e.g.*, a maximum running time, a minimum size of the sample, *etc.*), concluding the sampling process. Collected data are processed and stored in anonymized format,[a] post-processed, cleaned and filtered according to further requirements.

**The sampling methodologies**

In the following, we briefly discuss the two statistical sampling methods adopted in this work, namely the *breadth-first-search* and the *uniform* sampling.

*The breadth-first-search sampling*

The first adopted sampling methodology is a snowball technique that exploits the breadth-first-search (BFS), an uninformed graph traversal algorithm. Starting from a *seed node*, the procedure explores its neighborhood; then, for each neighbor, it visits its unexplored neighbors, and so on, until the whole network is visited (or, alternatively, a termination criterion is met). This sampling technique has several advantages with respect to other techniques (for example, *random walks* sampling, *forest fire* sampling, *etc.*) as discussed in recent literature [41, 42]. One of the main advantages is that it produces a coherent graph whose topological features can be studied.

For this reason it has been adopted in a variety of OSNs mining studies [1, 43–46]. During our experimentation, we defined the termination criterion that the mining process did not exceed 10 days of running time. By observing a short time-limit, we ensured a negligible effect of evolution of the network structure (less than 2% overall, according to a heuristic calculation based on the growth rate of Facebook during the sampling process - August 2010). The size of the obtained (partial) graph of the Facebook social network has been adopted as yardstick for the *uniform* sampling process.

*The uniform sampling*
The second chosen sampling methodology is a rejection-based sampling technique, called *uniform* sampling. The main advantage of this technique is that it is proven unbiased, at least in its formulation for Facebook. Details about its definition are provided by Gjoka *et al.* [44]. The process consists of generating an arbitrary number of user-IDs, randomly distributed in the domain of assignment of the Facebook user-ID system. In our case, it is the space of the 32-bit numbers: the maximum amount of assignable user-IDs is $2^{32}$, about 4 billions. As of August 2010 (the period during which we carried out the sampling process), the number of subscribed users on Facebook was about 500 millions, thus the probability of randomly generating an existing user-ID was $\approx 1/8$.

The sampling process has been set up as follows: first we generated a number of random user-IDs, lying in the interval $[0, 2^{32} - 1]$, equal to the dimension of the BFS-sample multiplied by 8. Then, we queried Facebook for their existence. Our expectation was to obtain a sample of comparable dimensions with respect to the BFS-sample. Actually, we obtained a slightly smaller sample, due to the restrictive privacy settings imposed by some users, who configured their profile preventing the public accessibility of their friend-lists. The issue of the privacy has been investigated in our previous work [45].

**Description of the samples**
All the user-IDs contained in the samples have been anonymized using a 48-bit hashing functions [47], in order to hide references to users and their connections. Data have been post-processed for a cleansing step, during which all the duplicates have been removed, and their integrity and congruency have been verified. The characteristics of the samples are reported in Table 1. The size of both the samples is in the magnitude of few millions of nodes and edges.

The anonymized datasets studied in this work may be examined by the scientific community.[b]

Some of the statistical and topological features of these networks have been discussed in our previous work [45], and our main previous findings can be summarized as follows:

- It emerges that the degree distribution of nodes in the samples is defined by a power law $P(x) \propto x^{-\lambda}$ identifying two different regimes. In detail, it is possible to divide the domain into two intervals (tentatively $1 \leq x \leq 10$ and $x > 10$), whose exponents are $\lambda_1^{BFS} = 2.45$, $\lambda_2^{BFS} = 0.6$ and $\lambda_1^{UNI} = 2.91$, $\lambda_2^{UNI} = 0.2$ respectively for the BFS and the *uniform* sample, in agreement with recent studies by Facebook [18, 19].
- Concerning the diameter of the networks, the BFS sample shows a small diameter, in agreement with the *six-degrees of separation theory* [39], given the snowball nature of

**Table 1 BFS and *uniform* samples description**

| Feature | BFS | *Uniform* |
|---|---|---|
| No. visited users | 63.4K | 48.1K |
| No. discovered neighbors | 8.21M | 7.69M |
| No. total edges | 12.58M | 7.84M |
| Size largest connected component | 98.98% | 94.96% |
| Avg. degree (visited users) | 396.8 | 326.0 |
| 2nd largest eigenvalue | 68.93 | 23.63 |
| Effective diameter | 8.69 | 14.72 |
| Avg. clustering coefficient | $1.88 \cdot 10^{-2}$ | $1.40 \cdot 10^{-3}$ |
| Density | 0.626% | 0.678% |

In this table we report some statistics regarding the two samples, BFS and *uniform*, which have been collected during August 2010 from the Facebook social network.

the sampling algorithm, which produces a plausible graph; differently, the diameter is over-represented in the *uniform* sample, possibly because the largest connected component does not cover the whole network.

- Regarding the *clustering coefficient*, we observed that the average values for both the samples are very high, similarly as reported by other recent studies on OSNs [3, 44]. High clustering coefficient and small diameter provide a clue of the presence of the so-called *small world* effect [15, 39] in the Facebook social graph.

**Detecting communities**

Given the large size of our Facebook samples, most of the community detection algorithms existing in literature could not deal with it. In order to unveil the community structure of these networks we adopted two computationally efficient techniques: (i) *Label Propagation Algorithm* (LPA) [48], and (ii) *Fast Network Community Algorithm* (FNCA) [49].

In the following we discuss the main advantages given from their choice and their performance.

*Advantages and performance of chosen methods*

The problem of selecting a particular community detection algorithm is crucial if the aim is to unveil the community structure of a network. In fact, the choice of a given methodology could affect the outcome of the experiments. In particular, several algorithms depend on tuning specific parameters, such as the size of the communities in the given networks, and/or their number (for additional information see recent surveys on this wide topic [30–35]).

In this study, the purpose was to unveil the unknown community structure of our Facebook samples, and to do so we choose two different techniques which rely just on the topology of the network itself as guide to discover the community structure.

LPA (Label Propagation Algorithm) is an algorithm for community detection based on the paradigm of label propagation, a common strategy characterizing several machine learning algorithms. Its computational cost is near liner with respect to the size of the analyzed network. This computational efficiency makes it well suited for the discovery of communities in large networks, such as in our case. LPA only exploits the network structure as guide and does not follow any pre-defined objective function to maximize (differently from FNCA); in addition, it does not require any prior information about the communities, their number or their size.

**Table 2  Results on Facebook network samples**

| Algorithm | No. communities | Network modularity | Time (s) |
|---|---|---|---|
| *BFS (8.21M vertices, 12.58M edges)* | | | |
| FNCA | 50,156 | 0.6867 | $5.97 \cdot 10^4$ |
| LPA | 48,750 | 0.6963 | $2.27 \cdot 10^4$ |
| *Uniform (7.69M vertices, 7.84M edges)* | | | |
| FNCA | 40,700 | 0.9650 | $3.77 \cdot 10^4$ |
| LPA | 48,022 | 0.9749 | $2.32 \cdot 10^4$ |

This table summarizes performance and results of the two chosen community detection algorithms (*i.e.*, FNCA and LPA) applied to the samples we collected from Facebook.

**Table 3  Representation of a community structure**

| Community-ID | List of members |
|---|---|
| community-$ID_1$ | {user-$ID_a$; user-$ID_b$; … ; user-$ID_c$} |
| community-$ID_2$ | {user-$ID_i$; user-$ID_j$; … ; user-$ID_k$} |
| … | {…} |
| community-$ID_N$ | {user-$ID_x$; user-$ID_y$; … ; user-$ID_z$} |

To represent the community structure discovered in each sample we adopted the format reported in this table.

FNCA (Fast Network Community Algorithm) is a computationally efficient method to unveil the community structure of large networks. It is based on the maximization of an objective function called *network modularity* [50, 51]. Similarly to LPA, it does not require prior information on the structure of the network, the number of communities present in the network and/or their size.

Even though the paradigms on which the algorithms rely are different, a common feature emerges: their functioning is agnostic with respect to the characteristics of the considered network. This aspect makes them an ideal choice, considering that we do not have any prior information about the characteristics of the community structure of Facebook. Further technical details regarding these methods are discussed in the Appendix of this paper.

The performance of the LPA and FNCA on our Facebook samples is showed in Table 2. Both the algorithms are able to unveil the community structure of the network in double time. High values of *network modularity* have been obtained in both the samples. This aspect suggests the presence of a well-defined community structure.

The community structure has been represented by using a list of vectors which are identified by a 'community-ID'; each vector contains the list of user-IDs (in anonymized format) of the users belonging to the given community; an example is depicted in Table 3. This representation was instrumental to carry out with efficiency the experiments discussed in the remainder of the paper.

*Assessing the quality of the community detection*

Remarkably, one challenge arises in the context of the assessment of the quality of the community detection process in real-world scenarios that is the lack of a ground truth against which to compare the results provided by the adopted community detection strategy - which might be biased by the strategy itself [37, 38]. For such a reason, some works [52–54] estimate the quality of a community detection algorithm by measuring some internal measures of quality of detected communities, based on topological characteristics (for example, the network modularity to establish the density of connections among and within communities, or the stability of eigenvalues of the Laplacian graph). Other approaches [33] are based on the possibility of exploiting exogenous factors, such as semantic information derived from additional knowledge on users (for example their affiliations to particular groups, *etc.*). In the first case, indicators of internal quality of the communities are often not sufficient to ensure the quality of the results - think, for example, at the resolution limit that arises in modularity maximization algorithms [37, 38]. On the other hand, no additional information on users other than their interconnections was available to us, for the purpose of assessing the quality of our communities.

Then, to establish the significance of the community structure obtained by using the methods discussed above, we chose to evaluate the similarity of outcomes provided by the two adopted algorithms, against each other, in a number of different ways which are discussed in the next section. This might help in highlighting anomalies in our methodology, in case of significant divergences between obtained results.

**Building the community meta-network**

To study the mesoscopic features of the community structure of Facebook, we abstracted a *meta-network* consisting of the communities, as follows. We built a weighted undirected graph $G' = (V', E', \omega)$, whose set of nodes is represented by the communities constituting the given community structure. In $G'$ there exists an edge $e'_{uv} \in E'$ connecting a pair of nodes $u, v \in V'$ if and only if there exists in the social network graph $G = (V, E)$ at least one edge $e_{ij} \in E$ which connects a pairs of nodes $i, j \in V$, such that $i \in u$ and $j \in v$ (*i.e.*, user $i$ belongs to community $u$ and user $v$ belongs to community $j$). The weight function is defined as

$$\omega_{u,v} = \sum_{i \in u, j \in v} e_{ij} \qquad (1)$$

(*i.e.*, the sum of the total number of edges connecting all users belonging to $u$ and $v$).

Table 4 summarizes some characteristics of the networks obtained for the *uniform* sample by using FNCA and LPA. Something which immediately emerges is that the overall statistics obtained by using the two different community detection methods are very similar. The number of nodes in the *meta-networks* is smaller than the total number of communities discovered by the algorithms, because we excluded all those 'communities' containing only one member (whose consideration would be in antithesis with the definition of community in the common sense).

**Table 4 Features of the meta-networks representing the *community structure* for the *uniform* sample**

| Feature | FNCA | LPA |
| --- | --- | --- |
| No. nodes/edges | 36,248/836,130 | 35,276/785,751 |
| Min./Max./Avg. weight | 1/16,088/1.47 | 1/7,712/1.47 |
| Size largest conn. comp. | 99.76% | 99.75% |
| Avg. degree | 46.13 | 44.54 |
| 2nd largest eigenvalue | 171.54 | 23.63 |
| Effective diameter | 4.85 | 4.45 |
| Avg. clustering coefficient | 0.1236 | 0.1318 |
| Density | 0.127% | 0.126% |

In this table we report some statistics regarding the community structure *meta-network* obtained from the *uniform* sample, by using the two chosen community detection algorithms (*i.e.*, FNCA and LPA).

We discuss results regarding the community structure and its mesoscopic features in the following.

## Results

The analysis of the community structure of Facebook will focus on the following aspects: (i) first, we try to evaluate the quality of the communities identified by means of the community detection algorithms described above. This step includes assessing the similarity of results obtained by using different sampling techniques and clustering methods. In detail, we evaluate the possible bias introduced by well-known limitations of these techniques (*e.g.*, the resolution limit for modularity maximization methods [37, 38] or the sampling bias due to the incompleteness of the sampling process [36]). (ii) Second, we investigate the mesoscopic features of the community *meta-network* considering some characteristics of the network (such as the diameter, the distribution of shortest-paths and weights of links, the connectivity among communities, *etc.*), discussing how these features may reflect organization patterns of individuals in the network.

### Analysis of the community structure

In order to characterize the features of the community structure of Facebook, our first step was to describe the distribution of the size of the communities discovered. This feature has been investigated in current literature [12, 40], and it emerges that different complex networks exhibit heavy-tailed distributions in the size of the communities. This implies the existence of a large amount of communities whose size is very small and a very small amount of large communities in this type of real-world networks. In detail, Lancichinetti *et al.* [12] put into evidence that this holds true for a large family of complex networks, such as information, communication, biological, and social networks.

*Distribution of the community size*

Figures 1 and 2 represent the probability mass function of the distributions of the size of discovered communities, respectively for *uniform* and BFS sample, by using the two cho-sen community detection algorithms. From the analysis of these figures, it emerges that in both cases the distributions produced by the two community detection algorithms are very similar. Moreover, we can observe that these distributions are broad and resembles other real-world complex networks (*cf.* [12]).

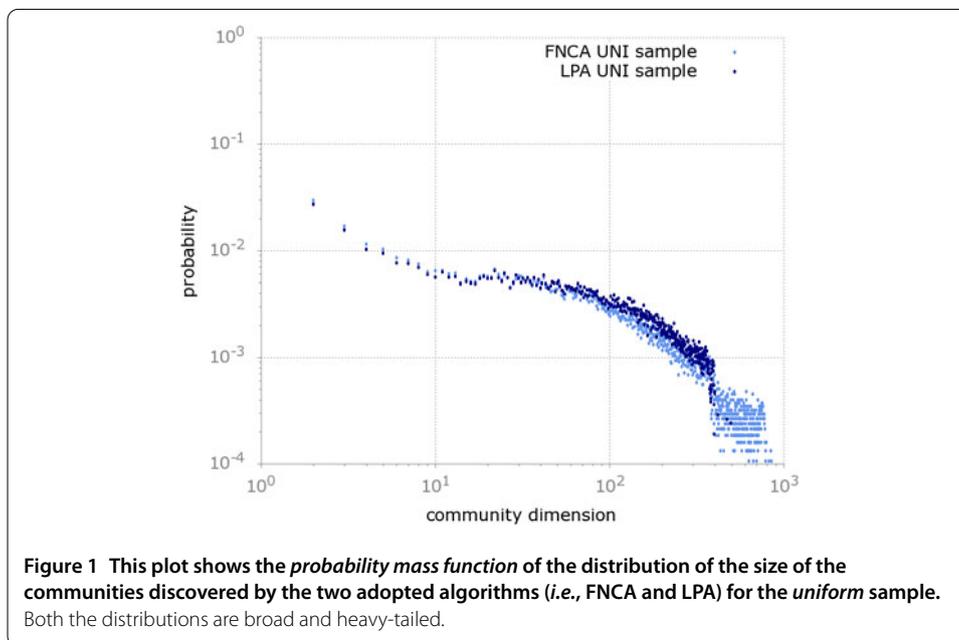

**Figure 1 This plot shows the *probability mass function* of the distribution of the size of the communities discovered by the two adopted algorithms (*i.e.*, FNCA and LPA) for the *uniform* sample.** Both the distributions are broad and heavy-tailed.

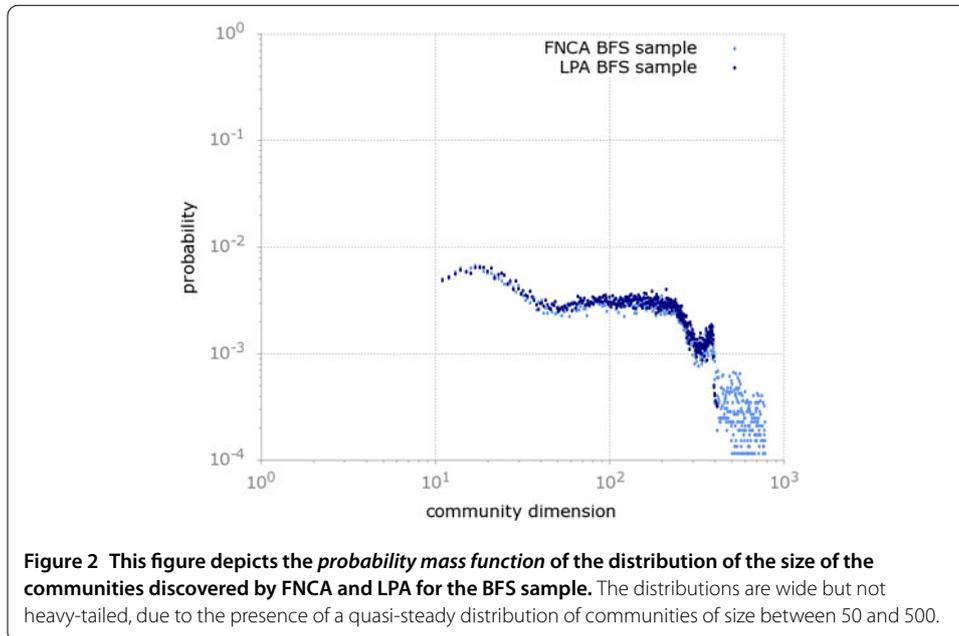

**Figure 2  This figure depicts the *probability mass function* of the distribution of the size of the communities discovered by FNCA and LPA for the BFS sample.** The distributions are wide but not heavy-tailed, due to the presence of a quasi-steady distribution of communities of size between 50 and 500.

From a further analysis it emerges that, for the *uniform* sample (Figure 1), both the distributions are broad and heavy-tailed. Differently, the distributions for the BFS sample are wide but not heavy-tailed, due to the quasi-steady probability of finding communities of size between 50 and 200.

The difference between BFS and *uniform* samples appears in agreement with the adopted sampling techniques. In fact, it has been recently put into evidence [36, 45] that a sampling algorithm such as the BFS may affect the degree distribution towards high degree nodes, in case of incomplete visits. Interestingly, this reflects also in the presence of communities, tentatively lying in the size interval $50 \geq x \geq 200$, that are in greater number with respect to what it would be expected by a scale-free network.

To the best of our knowledge, this is the first time it is observed that the bias towards high degree nodes introduced by the BFS sampling method reflects on the features of the community structure of a network. To the purpose of sampling, we could indicate as more appropriate those rejection-based methods, such as the *uniform* sampling, that do not over-represent high degree nodes.

Indeed, the analytical results reported in Table 2 combined with the plots discussed above, suggest that both the algorithms identified a similar amount of communities, regardless the adopted sampling method. This is also reflected by the similar values of *network modularity* obtained for the two different sets. Moreover, the size of the communities themselves seems to coincide for most of the times.

The following point we address is inspecting the quality of the community structure obtained by using FNCA and LPA. The possibility that two different algorithms produce different community structures is not to be excluded, thus in the following we investigate to what extent the results we obtained share a high degree of similarity.

*Community structure similarity*

In order to evaluate the similarity of two community structures we adopt three measures: (i) a variant of the *Jaccard coefficient*, called *binary Jaccard coefficient*; (ii) the *Kullback-Leibler divergence*; and, (iii) the *normalized mutual information*. In the following we discuss them separately, to explain their functioning, the motivations of their adoption and the obtained results.

The first measure considered to our purpose is the *binary Jaccard coefficient*, defined as

$$\hat{J}(\mathbf{v},\mathbf{w}) = \frac{M_{11}}{M_{01} + M_{10} + M_{11}}, \qquad (2)$$

where $M_{11}$ represents the total number of shared elements between two vectors[c] $\mathbf{v}$ and $\mathbf{w}$, $M_{01}$ represents the total number of elements belonging to $\mathbf{w}$ and not belonging to $\mathbf{v}$, and, finally $M_{10}$ the *vice-versa*. The outcome of this measure lies in [0,1].

The adoption of the binary Jaccard coefficient is due to the following consideration: if we would compute the simple intersection of two sets (*i.e.*, the community structures) by using the classic Jaccard coefficient, those communities differing even by only one member would be considered different, while a high degree of similarity among them could still be envisaged. We avoid this issue adopting the binary Jaccard coefficient, by comparing each vector of the former set against all the vectors in the latter set, in order to *match* the most similar one. The mean degree of similarity is then computed as

$$\sum_{i=1}^{N} \frac{\max(\hat{J}(\mathbf{v},\mathbf{w})_i)}{N}, \qquad (3)$$

where $\max(\hat{J}(\mathbf{v},\mathbf{w})_i)$ represents the highest value of similarity chosen among those calculated combining the vector $i$ of the former set with all the vectors of the latter set. We obtained the results as in Table 5, in which we show the mean, median and standard deviations of the results obtained by comparing, both for the BFS and the *uniform* sample, the outcome of the clustering processes according to the two different algorithms (*i.e.*, FNCA and LPA).

While the number of identical communities between the two sets obtained by using, respectively, BFS and *uniform* sampling, is not high (*i.e.*, respectively, $\approx 2\%$ and $\approx 35\%$), the overall mean degree of similarity is very high (*i.e.*, $\approx 73\%$ and $\approx 91\%$). This is due to the high number of communities which differ only for a very small number of elements. Moreover, the fact that the median is, respectively, $\approx 75\%$ and $\approx 99\%$, and that the very majority of results lie in one standard deviation, supports the similarity of the obtained community structures.

**Table 5 Similarity degree of community structures**

| Metric | Sample | Degree of similarity FNCA *vs.* LPA | | | |
|---|---|---|---|---|---|
| | | Common | Mean | Median | Std. D. |
| $\hat{J}$ | BFS | 2.45% | 73.28% | 74.24% | 18.76% |
| | uniform | 35.57% | 91.53% | 98.63% | 15.98% |

In this table we report the results obtained computing the similarity between the community structure discovered by using FNCA and LPA in the BFS and *uniform* samples, computed by means of the binary Jaccard coefficient.

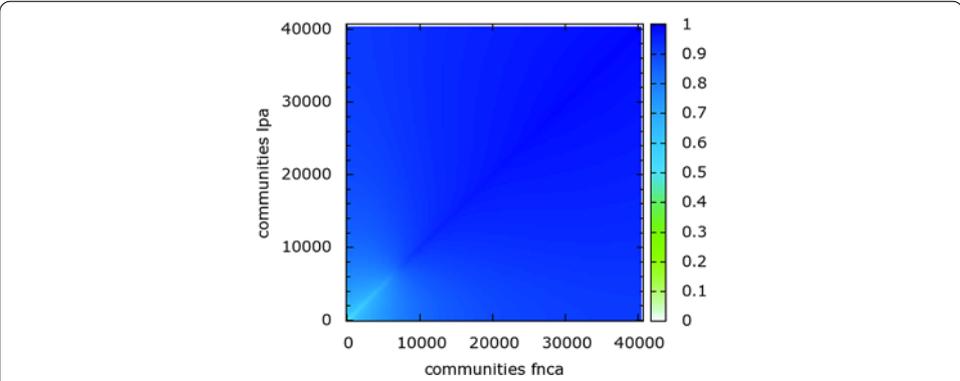

**Figure 3 This heat-map highlights the similarity of the communities discovered by means of the two adopted algorithms (*i.e.*, FNCA and LPA) applied to the *uniform* sample.** Almost the totality of communities discovered share a high fraction of members (in average the 91%), according the Jaccard similarity computed pairwisely selecting the most similar communities in the partitions.

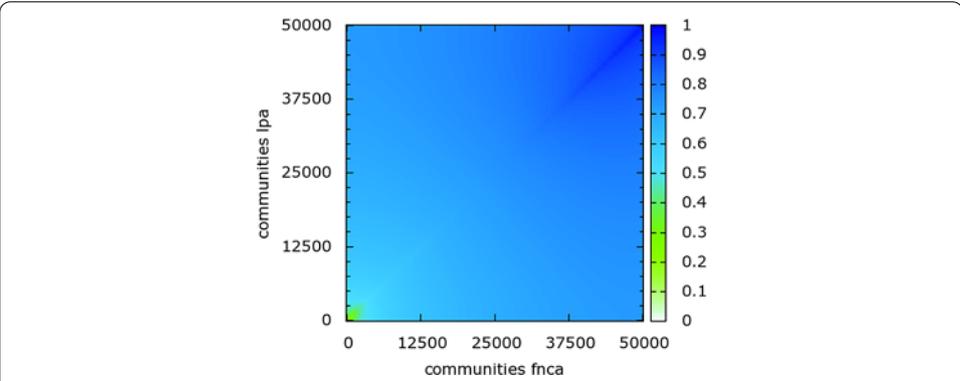

**Figure 4 This heat-map shows the similarity of the communities discovered by FNCA and LPA in the BFS sample.** In this case, with respect to the *uniform* sample case, the pairwise similarity between communities emerges slightly less obviously, but it is in average the 73%.

Figures 3 and 4 graphically highlight these findings. Their interpretation is as follows: on the *x-axis* and on the *y-axis* there are represented the communities discovered for the FNCA and the LPA methods, respectively. The higher the degree of similarity between two compared communities, the higher the heat-map scores. The similarity is graphically evident considering that the values of heat showed in the figures are very high (*i.e.*, greater than 0.7) for the most of the heat-map.

Before introducing the second experiment, observe that it is desirable to assess, not only if the two clustering solutions present a large amount of similar clusters, but also if they exhibit a similar statistical distribution in the size of the obtained clusters. To this purpose, a second method has been taken into consideration: the divergence measure called *Kullback-Leibler divergence*, that is defined as

$$D_{\mathrm{KL}}(P \parallel Q) = \sum_i P(i) \log \frac{P(i)}{Q(i)}, \qquad (4)$$

where *P* and *Q* represent, respectively, the probability distributions that characterize the size of communities discovered by LPA and FNCA, calculated on a given sample. Let *i* be a given size such that $P(i)$ and $Q(i)$ represent the probability that a community of size $i$ exists in the distributions $P$ and $Q$. The KL divergence is helpful if one would like to calculate how different are two distributions with respect to one another. In particular, being the KL divergence defined in the interval $0 \leq D_{KL} \leq \infty$, the smaller the value of KL divergence between two distributions, the more similar they are.

We calculated the pairwise KL divergences between the distributions discussed above, finding the following results.

(i) on the *uniform* sample:
- $D_{KL}(P_{LPA} \| P_{FNCA}) = 7.722 \cdot 10^{-3}$
- $D_{KL}(P_{FNCA} \| P_{LPA}) = 7.542 \cdot 10^{-3}$

(ii) on the BFS sample:
- $D_{KL}(P_{LPA} \| P_{FNCA}) = 3.764 \cdot 10^{-3}$
- $D_{KL}(P_{FNCA} \| P_{LPA}) = 4.292 \cdot 10^{-3}$

The low values obtained by adopting the KL divergence put into evidence the correlation between the distributions calculated by using the two different algorithms on the two different samples.

Finally, to compute the quality of the results, we adopted a third measure, called *normalized mutual information* (NMI) [55]. Such a measure assumes that, given a graph $G$, a *ground truth* is available to verify what are the clusters (said *real clusters*) in $G$ and what are their features. Let us denote as $A$ the true community structure of $G$ and suppose that $G$ consists of $c_A$ clusters. Let us consider a clustering algorithm applied on $G$ and assume that it identifies a community structure $B$ consisting of $c_B$ clusters. We define a $c_A \times c_B$ matrix - said *confusion matrix* - $CM$ such that each row of $CM$ corresponds to a cluster in $A$ whereas each column of $CM$ is associated with a cluster in $B$. The generic element $CM_{ij}$ is equal to the number of elements of the real $i$th cluster which are also present in the $j$th cluster found by the algorithm. Starting from these assumptions, the *normalized mutual information* is defined as

$$\text{NMI}(A,B) = \frac{-2 \sum_{i=1}^{c_A} \sum_{j=1}^{c_B} N_{ij} \log(\frac{N_{ij} N}{N_{i.} N_{.j}})}{\sum_{i=1}^{c_A} N_{i.} \log(\frac{N_{i.}}{N}) + \sum_{j=1}^{c_B} N_{.j} \log(\frac{N_{.j}}{N})} \tag{5}$$

being $N_{i.}$ (resp., $N_{.j}$) the sum of the elements in the $i$th row (resp., $j$th column) of the confusion matrix. If the considered clustering algorithm would work perfectly, then for each discovered cluster $j$, it would exist a real cluster $i$ exactly coinciding with $j$. In such a case, it is possible to show that $\text{NMI}(A,B)$ is exactly equal to 1 [55]. By contrast, if the clusters detected by the algorithm are totally independent of the real communities then it is possible to show that the NMI is equal to 0. The NMI, therefore, ranges from 0 to 1 and the higher the value, the better the clustering algorithm performs with respect to the ground truth.

Observe that, in our scenario, we do not have at disposal any ground truth, given the fact that the community structure of the considered network is unknown and the purpose of our study was, in fact, to discover it. Still, we can adopt the NMI to compare against each other, the outcome of the clustering solutions obtained by means of two different algorithms, taking the result of the former and using it as a ground truth to assess the latter, or the *vice-versa*.

Several variants of *normalized mutual information* exist: to our purposes, we adopted two different versions of NMI, henceforth called $NMI_{LFK}$ and $NMI_{MGH}$ - after the authors initials - presented, respectively, by Lancichinetti, Fortunato and Kertesz [56] and by McDaid, Greene and Hurley [57]. These two variants adopt slightly different normalization factors, thus they produce different (but comparable) results.

Applying these two versions of NMI according to the considerations presented above, we obtained the following results:

(i) on the *uniform* sample:
- $NMI_{LFK}(FNCA_{uniform}, LPA_{uniform}) = 0.825$
- $NMI_{MGH}(FNCA_{uniform}, LPA_{uniform}) = 0.786$

(ii) on the BFS sample:
- $NMI_{LFK}(FNCA_{BFS}, LPA_{BFS}) = 0.678$
- $NMI_{MGH}(FNCA_{BFS}, LPA_{BFS}) = 0.648$

The high values obtained by using the *normalized mutual information*, which is able to better capture nuances and facets of different clustering solutions with respect to the much simpler binary Jaccard coefficient adopted above, still confirm the similarity of the community structure discovered by the two different algorithms employed in our analysis.

Given the limitations imposed by the lack of a ground truth for real-world networks for which the community structure is unknown, the approaches we adopted to assess the results are only a best-approximation of any robust evaluation method. Indeed, the problem of evaluating the clustering quality of real-world networks lacking of a ground truth is an open and urgent problem in current literature.

In addition, recently [37], in the context of detecting communities by adopting the *network modularity* as maximization function, a resolution limit has been put into evidence. In [37], the authors found that modularity optimization could, depending on the topology of the network, cause the inability of the process of community detection to find communities whose size is smaller than $\sqrt{E/2}$ (*i.e.*, in our case $\approx 3,000$). This reflects in another effect, that is the creation of big communities that include a large part of the nodes of the network, without affecting the global value of network modularity.

Being all the communities revealed smaller than that size and distributed in agreement with what already observed for other complex networks [12], we may hypothesize that the community structure unveiled by the algorithm for our samples is unlikely to be affected by the resolution limit.

**Mesoscopic features of the community structure**

In the following we consider the *uniform* sample and the community structure unveiled by the LPA as yardstick for our investigation. The experiments discussed in the remainder of this section focus in particular on three aspects: (i) assessment of the mesoscopic features of the community structure of the network and their implications in terms of social dynamics; (ii) study of the connectivity among communities and how it reflects on users organization patterns on a large scale; (iii) ability of inferring additional insights by means of visual observation of the community structure.

The purpose of investigating the mesoscopic features of the community structure of Facebook includes finding patterns that emerge from the network structure, in particular those which are related not to individuals or to the overall networks, but with those aggregation units that are the communities among which users gather.

To this purpose, we first discuss the degree distribution of communities discovered by means of our methods (*i.e.*, FNCA and LPA) in the *uniform* sample. We report Figure 5, that shows the complementary cumulative degree probability distribution (ccdf) as a function of the degree in the cases discussed above. The meaning of the *complementary cumulative distribution function (ccdf)*, defined as $F(x) = \Pr(X > x)$, is the probability that a random variable $X$ assumes values below a given $x$. Analyzing these distributions we observe a very peculiar feature: two different regimes, tentatively $1 \leq x < 100$ and $100 \leq x < 300$, can be identified, and a cut-off in proximity of $x \approx 300$ as well. This reveals a decreasing chance of finding communities as their size grows, with a clear cut-off above a certain threshold. Interestingly, a similar phenomenon has been previously observed in the Facebook social graph [44] and it has been put in correlation with the so-called *self-organization* principle observed in social networks [58]. Self-organization is the ability of individual to coordinate and organize in patterns or structures which are proven to be efficient, robust and reliable. For example, efficiency could be expressed in terms of minimizing costs for diffusing information [59, 60], robustness could be represented by the presence of redundant connections that link the same groups and reliability by the ability of the network to well-react to errors and malfunctioning [61–63].

In the light of this observations, we tried to relate how communities grow with respect to their degree of connectivity. Our finding are reported in Figure 6. It emerges that, not only the communities above a certain threshold size are much less likely to happen, but also they are much less interconnected. In fact, we can observe that the average degree of communities grows proportionally to their size up to a cut-off value still approximately $x \approx 300$. Above this threshold, larger communities become less and less connected with the others. This finding provides an argument in support to the idea that individuals in online social networks are mostly aggregated in small- or medium-size communities. On the other hand, large communities may suffer of a lack of external connectivity. The fact that individuals mostly aggregate in communities well-connected among each other without a coordinated effort is in line with the self-organization principle explained above.

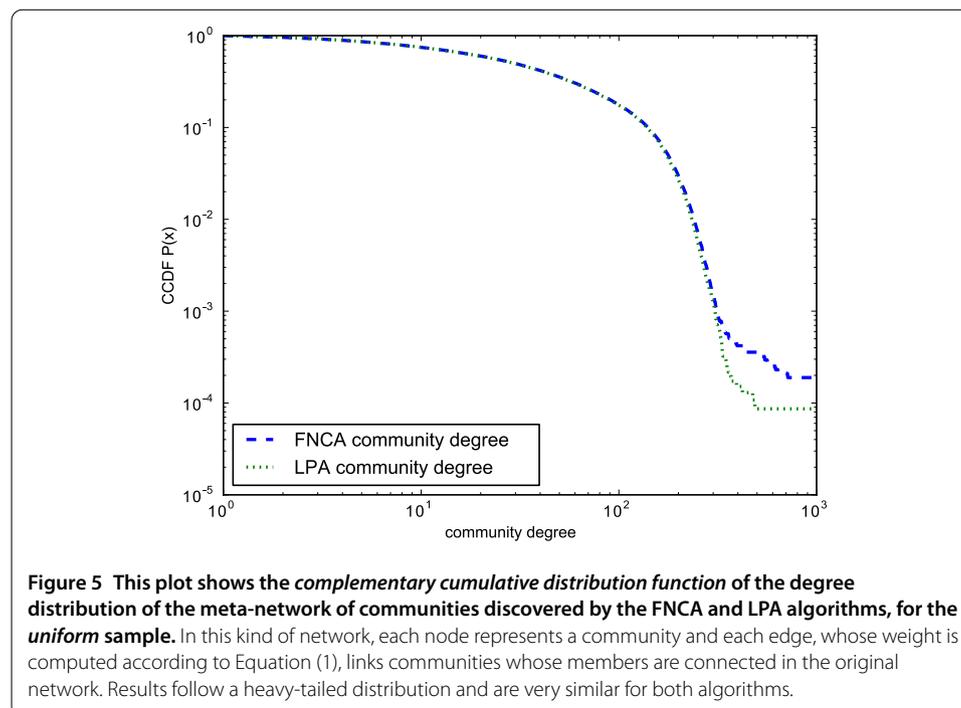

**Figure 5 This plot shows the *complementary cumulative distribution function* of the degree distribution of the meta-network of communities discovered by the FNCA and LPA algorithms, for the *uniform* sample.** In this kind of network, each node represents a community and each edge, whose weight is computed according to Equation (1), links communities whose members are connected in the original network. Results follow a heavy-tailed distribution and are very similar for both algorithms.

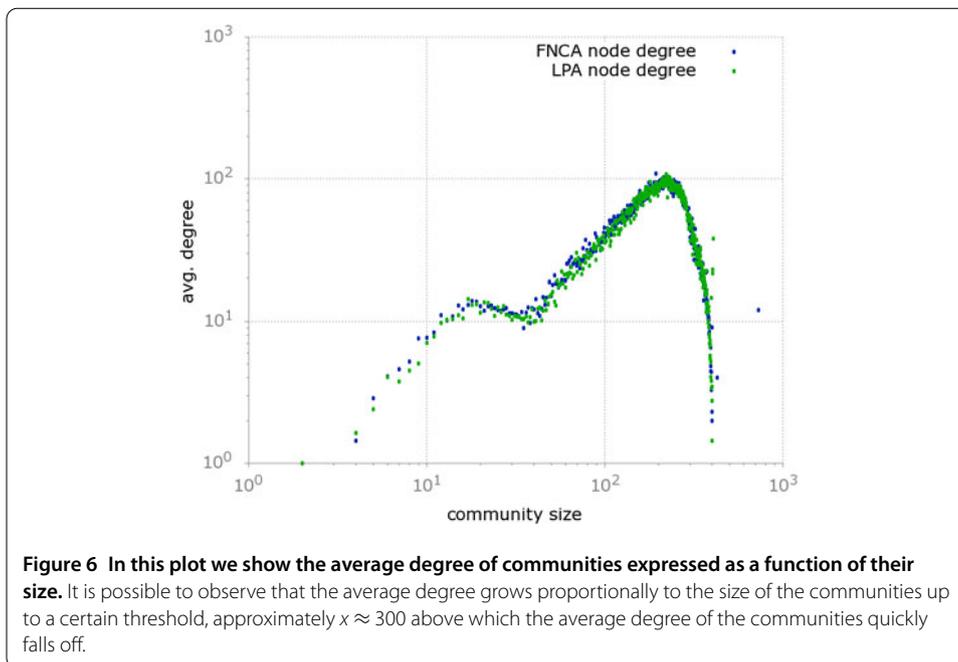

**Figure 6 In this plot we show the average degree of communities expressed as a function of their size.** It is possible to observe that the average degree grows proportionally to the size of the communities up to a certain threshold, approximately $x \approx 300$ above which the average degree of the communities quickly falls off.

Interestingly, self-organization is a phenomenon which is known to happen in *small world* networks [59, 60, 64, 65] and in their community structure [66]. In the light of this assumption, we investigated the presence of the *small world* effect in the community structure of Facebook. To this purpose, a reliable indicator of the presence of this phenomenon is the clustering coefficient - *i.e.*, the tendency to the creation of closed triangles among triads of communities. In our context, the clustering coefficient of a community is the ratio of the number of existing links over the number of possible links between the given community and its neighbors. Given our meta-network $G = (V, E)$, the clustering coefficient $C_i$ of community $i \in V$ is

$$C_i = 2\left|\left\{(v, w) | (i, v), (i, w), (v, w) \in E\right\}\right| / k_i(k_i - 1),$$

where $k_i$ is the degree of community $i$.

It can be intuitively interpreted as the probability that, given two randomly chosen communities that share a common neighbor, there also exists a link between them. High values of average clustering coefficient indicate that the communities are well connected among each other. This result would be interesting since it would indicate a tendency to the *small world* effect.

We plotted the average clustering coefficient probability distribution for the community structure in Figure 7. From its analysis it emerges that the slope of this curve is smooth, which allows for a the existence of a high probability of finding communities with large clustering coefficient, irrespectively of the number of connections they have with other communities.

This interesting fact reflects the existence of a tight and highly connected core in the community structure [25, 26]. The *small world* effect is also related to the presence of short-paths connecting communities. In this context, it is reasonable to suppose that, randomly selecting two disconnected communities, it is likely that a short path connecting their members exists.

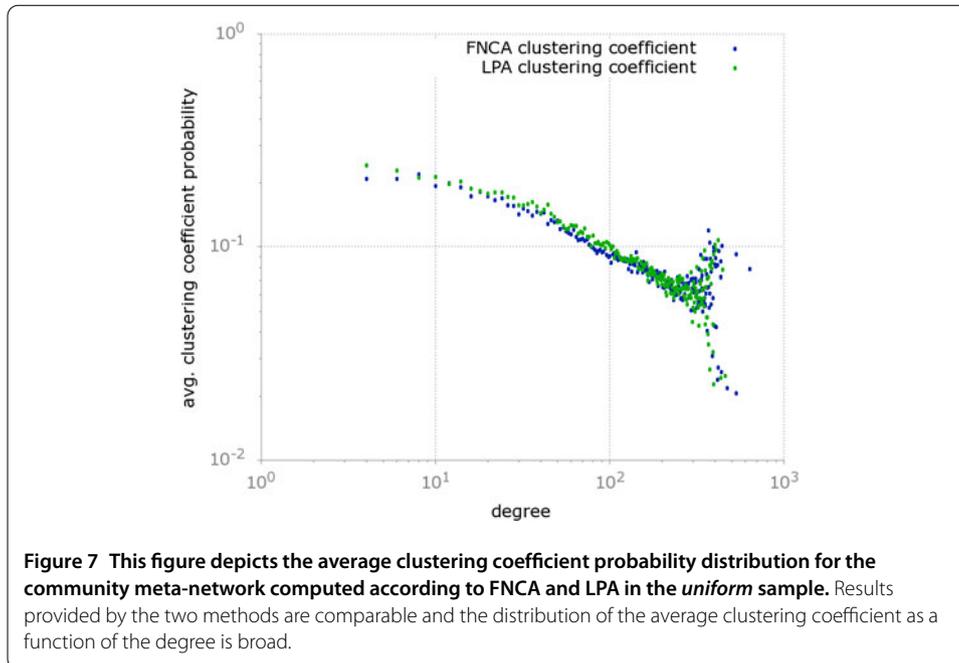

**Figure 7 This figure depicts the average clustering coefficient probability distribution for the community meta-network computed according to FNCA and LPA in the *uniform* sample.** Results provided by the two methods are comparable and the distribution of the average clustering coefficient as a function of the degree is broad.

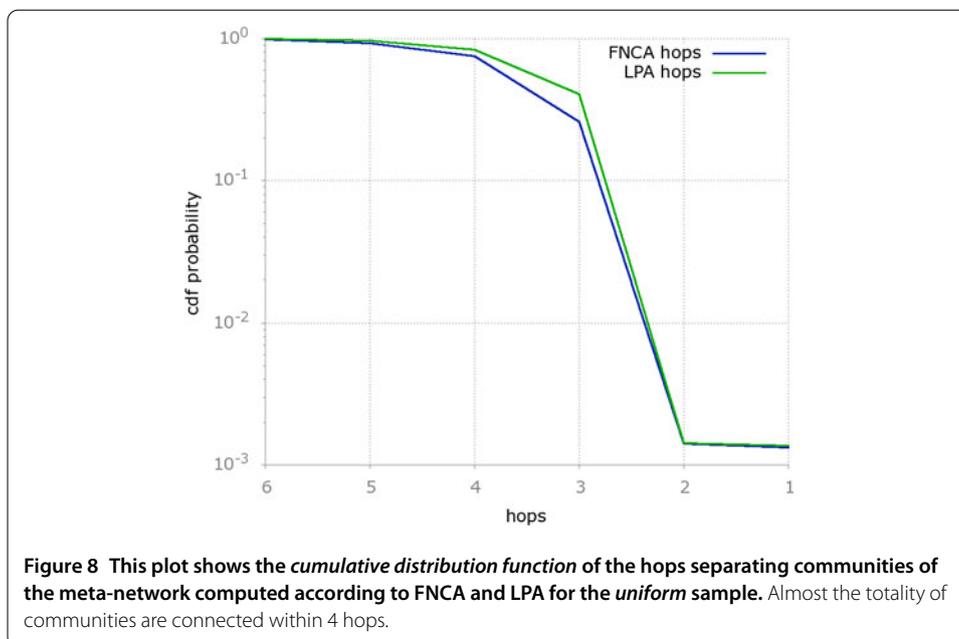

**Figure 8 This plot shows the *cumulative distribution function* of the hops separating communities of the meta-network computed according to FNCA and LPA for the *uniform* sample.** Almost the totality of communities are connected within 4 hops.

To investigate this aspect, in the following we analyze the effective diameter and the shortest paths distribution in the community structure. To this purpose, Figure 8 reports the *cumulative distribution function* of the probability that two arbitrary communities are connected in a given number of hops. The *cumulative distribution function (cdf)* defines the probability that a random variable $X$ assumes values below a given $x$. In that sense, from Figure 8 it emerges[d] that all communities are connected in a number of hops of 6, and most interestingly, that the highest advantage in terms of probability gain of connecting two randomly chosen communities, is obtained considering hops of length 3.

This aspect is further investigated as follows: Figure 9 represents the probability distribution for the shortest paths as a function of the path length. The interesting behavior which emerges from its analysis is that the shortest path probability distribution reaches a peak for paths of length 2 and 3. In correspondence with this peak, the number of connected pairs of communities quickly grows, reaching the effective diameter of the networks (*cf.* Figure 8). This findings has an important impact on the features of the overall social graph. In fact, if we would suppose that all nodes belonging to a given community are well connected each other, or even directly connected, this would result in a very short diameter of the social graph itself. In fact, there will always exist a very short path connecting the communities of any pair of randomly chosen members of the social network. Interestingly, this hypothesis is substantiated by recent studies by Facebook, who used heuristic techniques to measure the average diameter of the whole network [18, 19]. Their outcomes are very similar to our results: they estimated an average diameter of 4.72 while the effective diameter of the community structure for our *uniform* sample is 4.45 and 4.85, respectively for LPA and FNCA.

Thus, we conclude the characterization of the mesoscopic features of the community structure discussing the distribution of weights and strength of links among communities. The importance of this kind of analysis rises considering some social conjectures, like the Granovetter's *strength of weak ties theory* [24], that rely on the assessment of the strength of links in social networks. To this purpose, we resemble that the *strength* $s^\omega(v)$ (or *weighted degree*) of a given node $v$ is determined as the sum of the weights of all edges incident on $v$, defined as

$$s^\omega(v) = \sum_{e \in I(v)} \omega(e),$$

where $\omega(e)$ is the weight of a given edge $e$ and $I(v)$ the set of edges incident on $v$.

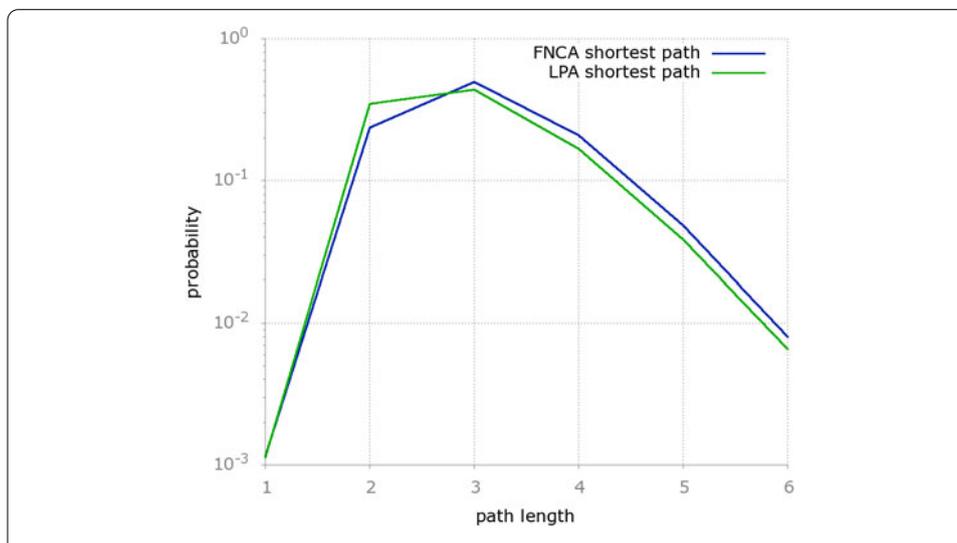

**Figure 9 This figure displays the shortest paths probability distribution for the meta-network of communities discovered by the FNCA and LPA algorithms in the *uniform* sample.** In agreement to Figure 8, the path length peak resides between 2 and 4.

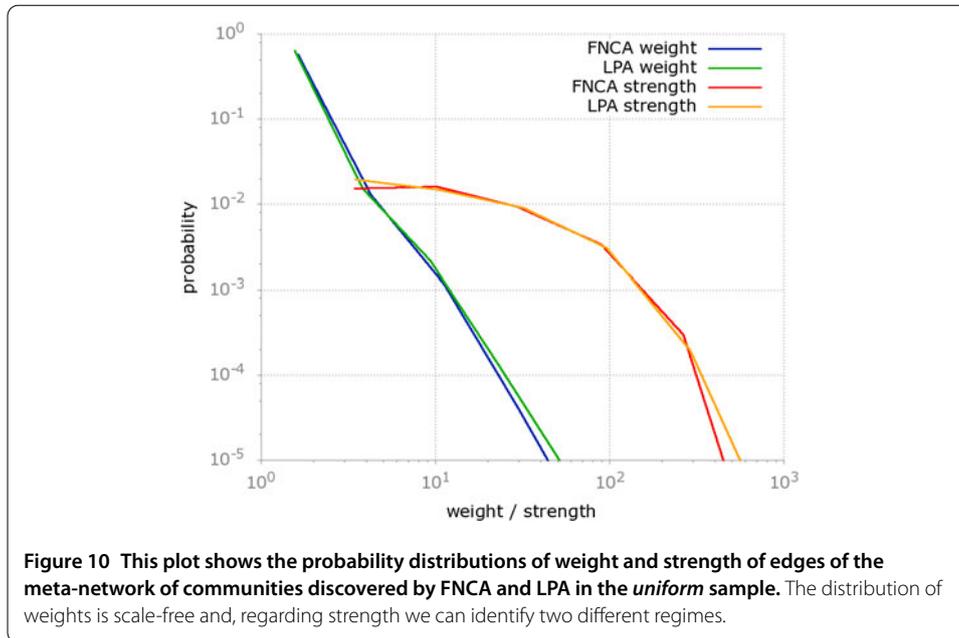

**Figure 10  This plot shows the probability distributions of weight and strength of edges of the meta-network of communities discovered by FNCA and LPA in the *uniform* sample.** The distribution of weights is scale-free and, regarding strength we can identify two different regimes.

In Figure 10, we plotted the probability distribution of both weight and strength on links among communities. Interestingly, the distribution of weights is defined by a power law $P(x) = x^{-\gamma}$ described by a coefficient $\gamma = 1.45$. The strength distribution is still broad but it is possible to observe two different regimes, in the intervals of tentatively $1 \leq x < 10^2$ and $x \geq 10^2$.

Given the definition of weights for the community meta-network, as in Equation (1) (*i.e.*, the sum of total number of edges connecting all users belonging to the two connected communities), we can suggest the hypothesis that there exists a high probability of finding a large number of pairs of communities whose members are not directly connected, and a increasingly smaller number of pairs of communities whose members are highly connected each other. These connections, which are usually referred as to *weak ties*, according to the *strength of weak ties theory* [24], are characterized by a smaller strength but a hightened tendency to proficiently connect communities otherwise disconnected. This aspect is further discussed in the following.

**Connectivity among communities**

The last experiment discussed in this paper is devoted to understanding the density of links connecting communities in Facebook. In particular, we are interested in defining to what extent links connect communities of comparable or different size. To do so, we considered each edge in the community *meta-network* and we computed the size of the community to which the *source node* of the edge belonged to. Similarly, we computed the size of the *target* community.[e]

Figure 11 represents a probability density map of the distribution of edges among communities. First, we highlight that the map is symmetric with respect to the diagonal, according to the fact that the graph is undirected and each edge is counted twice, once for each end-vertex. From the analysis of this figure, it emerges that edges mainly connect two types of communities: (i) communities of small size, each other - this is the most common case; (ii) communities of small size with communities of large size - less likely to happen but still significant.

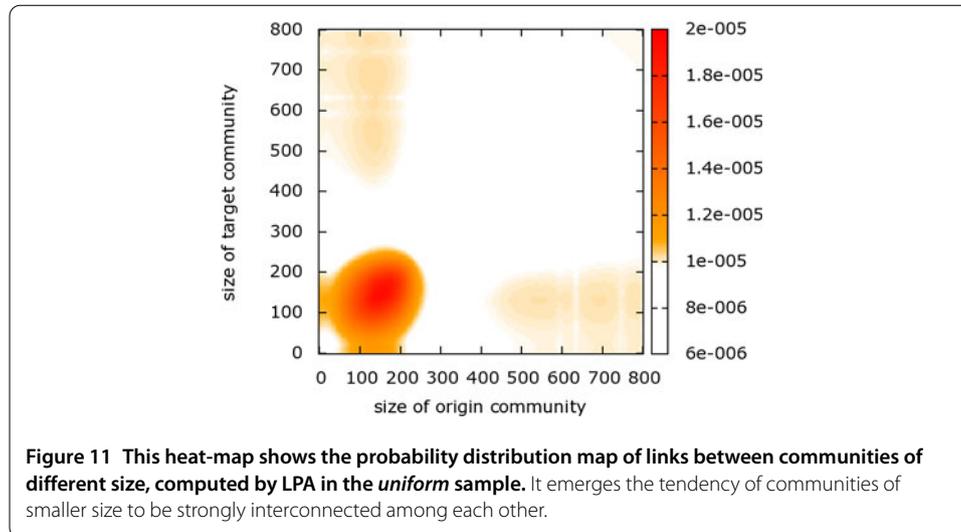

**Figure 11 This heat-map shows the probability distribution map of links between communities of different size, computed by LPA in the *uniform* sample.** It emerges the tendency of communities of smaller size to be strongly interconnected among each other.

This can be intuitively explained since the number of communities of small size is much greater than the number of large communities. On the other hand, it is an important finding since similar results have been recently described for Twitter [21], in the context of the evaluation of the Granovetter's *strength of weak ties theory* [24].[f]

In fact, according to this theory, weak links typically occur among communities that do not share a large amount of neighbors, and are important to keep the network proficiently connected.

*Inter and intra-community links*
For further analysis, we evaluated the amount of edges that fall in each given community with respect to its size. The results of this assessment are reported in Figure 12. The interpretation of this plot is the following: on the *y-axis* it is represented the fraction of edges per community as a function of the size of the community itself, reported on the *x-axis*. It emerges that also the distribution of the link fraction against the size of the communities follows a power law with an exponent equal to $x = 2.45$. This result shows that small communities are also more internally dense, while larger communities exhibit less internal connectivity - decreasing according to their size. Indeed, this result is different from that recently proved for Twitter [21], in which a Gaussian-like distribution has been discovered. This is probably due to the intrinsic characteristics of the networks, that are topologically dissimilar (*i.e.*, Twitter is represented by a directed graph with multiple type of edges) and also the interpretation itself of social tie is different. In fact, Twitter represents in a way *hierarchical connections -* in the form of *follower* and *followed* users - while Facebook tries to reflects a friendship social structure which better represents the community structure of real social networks.

The emergence of this scaling law is interesting with regard to the organization patterns that are reflected by individuals participating to large social networks. In fact, it seems that users that constitute small communities are generally very well connected to other communities and among each others, while large communities of individuals seem to be linked in a less efficient way to other communities - and also less dense of links. This is reflected by the small number of weak ties incident on communities of large size with respect to the number of individuals they gather. These findings are relevant since they provide a clue that individuals are able to self-organize even in large networks and without a coordinated effort. This might improve their ability to efficiently get in touch and communicate with a number of users larger than their friends or acquaintances.

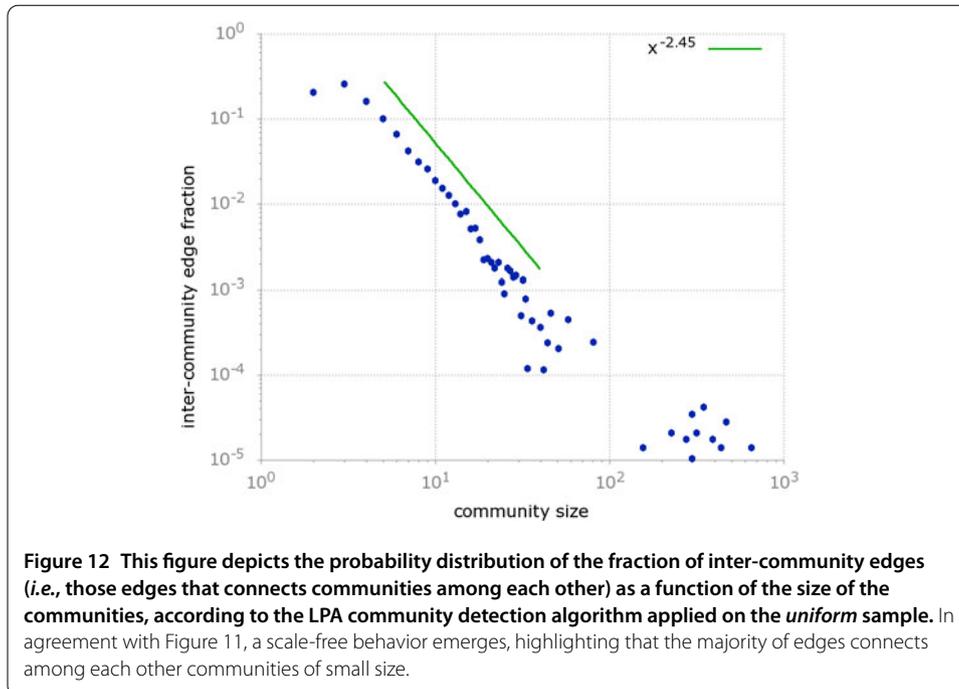

**Figure 12  This figure depicts the probability distribution of the fraction of inter-community edges (*i.e.*, those edges that connects communities among each other) as a function of the size of the communities, according to the LPA community detection algorithm applied on the *uniform* sample.** In agreement with Figure 11, a scale-free behavior emerges, highlighting that the majority of edges connects among each other communities of small size.

**Visual observation of the community meta-network**

The visual analysis of large-scale networks is usually unfeasible when managing samples whose size is in the order of millions of entities. Even though, by adopting our technique of building a community meta-network, it is yet possible to study the mesoscopic features of the Facebook social network from an unprecedented perspective. To this purpose, for example, social network analysts may be able to infer additional insights about the structure of the original network from the visual analysis of its community structure.

In Figure 13, obtained by using Cvis[g] - a hierarchical-based circular visualization algorithm - we represent the community structure unveiled by LPA in the *uniform* sample. From its analysis, it is possible to appreciate the existence of a tight core of communities which occupy a central position into the meta-network [25, 26]. A further inspection of the features of these communities revealed that their positioning is generally irrespective of their size. This means that there are several different small communities which play a dominant role in the network. This is in agreement with previous findings and highlight the role of self-organization on such a scale. Similar considerations hold for the periphery of the network, which is constituted both by small and larger communities.

Finally, we highlight the presence of so-called *weak ties*, that proficiently connect communities that otherwise would be far each other. In particular, those that connect communities in the core with communities in the periphery of the network, according to the *strength of weak ties theory* [24], might represent the most important patterns along which communications flow, enhancing users ability of getting in touch with each other, efficiently spreading information, and so on.

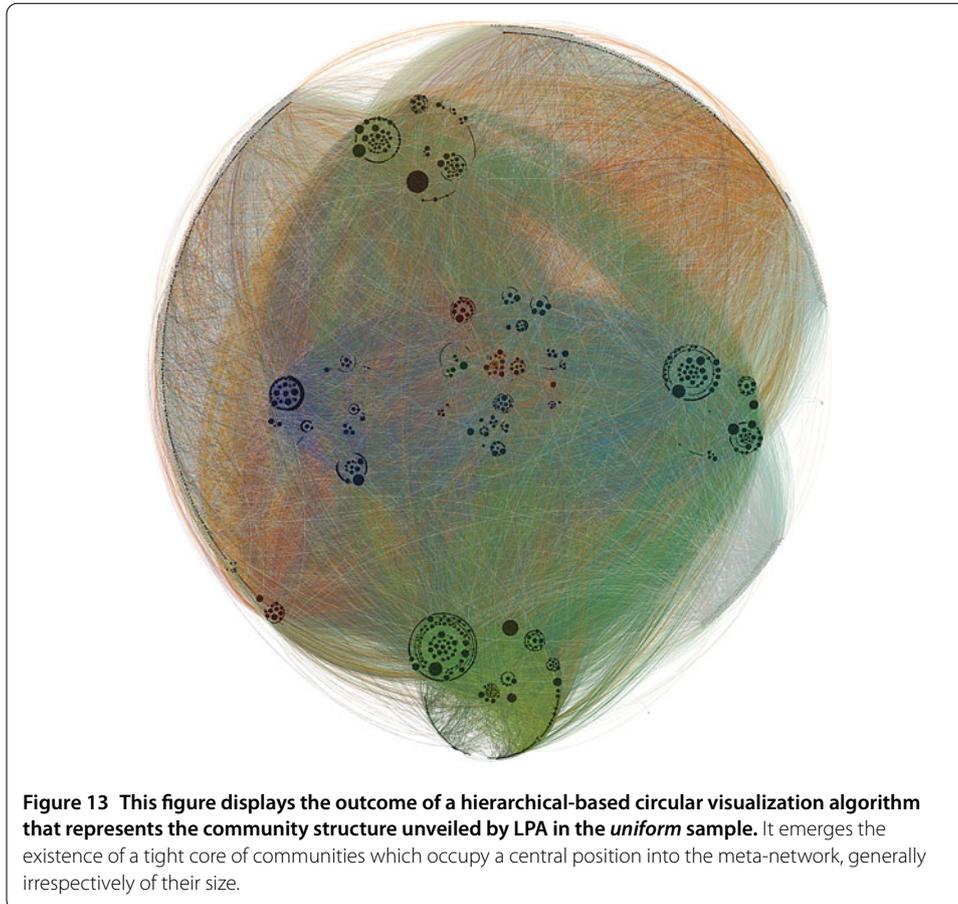

**Figure 13 This figure displays the outcome of a hierarchical-based circular visualization algorithm that represents the community structure unveiled by LPA in the *uniform* sample.** It emerges the existence of a tight core of communities which occupy a central position into the meta-network, generally irrespectively of their size.

## Discussion

This work concludes putting into evidence implications, strength and limitations of our study.

First of all, in this paper we put into evidence that the community structure of the Facebook social network presents a broad distribution of the dimension of the communities, similarly to other complex networks [12]. This result is independent with respect to the algorithm adopted to discover the community structure, and even (but in a less evident way) to the sampling methodology adopted to collect the samples. On the other hand, this is the first experimental work that proves the hypothesis, theoretically advanced by [36], of the possible bias towards high degree nodes introduced by the BFS sampling methodology for incomplete sampling of large networks.

Regarding the qualitative analysis of our results, it emerges that the communities share a high degree of similarity among different samples.

The analysis of the community meta-network puts into evidence different mesoscopic features. We discovered that the average degree of communities average degree of communities and their size put into evidence the tendency to self-organization of users into small- or medium-size communities well-connected among each other.

Our further analysis highlights that there exists a tendency to the creation of short-paths (whose length mainly consists of two or three hops), that proficiently connect the majority of the communities existing in the network. This finally led us to the identification of links connecting communities otherwise disconnected, that we called *weak ties* in the Granovetter's sense [24].

**Results in context with previous literature**

Several recent studies focused on the analysis of the community structure of different social networks [12, 13, 40, 67]. An in-depth analysis of the Facebook collegiate networks has been carried out in [13]. Authors considered data collected from 5 American colleges and examined how the online social lives reflect the real social structure. They proved that the analysis of the community structure of online social networks is fundamental to obtain additional insights about the prominent motivations which underly the community creation in the corresponding real world. Moreover, authors found that the Facebook social network shows a very tight community structure, and exhibits high values of network modularity. Some of their findings are confirmed in this study on a large scale.

Recently [12], it has been put into evidence that the community structure of social networks shares similarities with communication and biological networks. The authors investigated several mesoscopic features of different networks, such as community size distribution, density of communities and the average shortest path length, finding that these features are very characteristic of the network nature. According to their findings, we assessed that also Facebook is well-described by some specific characteristics on a mesoscopic level.

Regarding the mesoscale structure analysis of social networks, [40] provided a study by comparing three state-of-the-art methods to detect the community structure on large networks. An interesting aspect considered in that work is that two of the three considered methods can detect overlapping communities, so that a differential analysis has been carried out by the authors. They focused on the analysis of several mesoscopic features such as the community size and density distribution and the neighborhood overlapping. In addition, they verified that results obtained by the analysis of synthetic networks are profoundly different from those obtained by analyzing real-world datasets, in particular regarding the community structure, putting into evidence the emergence of need of studying online social networks acquiring data from the real platforms. Their findings are also confirmed in this study, in which we acquired a sample of the social graph directly from the Facebook platform.

An interesting work which is closely related to this study regards the assessment of the *strength of weak ties theory* in the context of Twitter [21]. In that work, it emerges that one of the roles of weak ties is to connect small communities of acquaintances which are not that close to belong to the same community but, on the other hand, are somehow proficiently in contact. Clues in this direction come also from this study, although the two networks exhibit different topological features (*i.e.,* Twitter is represented by a directed graph with multiple type of edges) and also carry a different interpretation of the social connections themselves. In fact, social ties in Twitter represent *hierarchical connections* (in the form of *follower* and *followed* users), while Facebook tries to reflects a friendship social structure which better represents the community structure of real-world social networks.

Concluding, recently [67] the perspective of the study of the community structure has been *revisited* considering the problem of the detecting communities of edges instead of the classical communities of nodes. In this approach we observe an interesting feature, *i.e.*, that link communities intrinsically incorporate the concept of overlap. The authors findings are applied to large social networks of mobile phone calls confirming the emergence of scale free distributions also for link community structures. Similar studies could be extended to online social networks like Facebook, in order to investigating the existence of particular communication patterns or motifs.

**Strength and limitations of this study**

In the following we discuss the main strengths and limitations of this study. To the best of our knowledge, this is the first work that investigates the general mesoscopic structure of a large online social network. This is particularly interesting since it is opposed to just trying to identify dense clusters in large communities, which is the aim of different works discussed above.

This work highlights the possibility of inferring characteristics describing the organization patterns of users of large social networks, analyzing some mesoscopic features that arise from a statistical and topological investigation. This kind of analysis has been recently carried out for some types of social media platforms (such as Twitter [21]) which capture different nuances of relations (for example, hierarchical follower-followed user relations), but there was a lack in literature regarding online social network platforms reflecting friendship relations, such as Facebook. This work, that tries to fill this gap, provides results that well relate with those presented in recent literature, and describes novel insights on the problem of characterizing social network structure on the large scale.

We can already envision two limitations of this work, which leave space for further investigation. First, our sample purely relies on binary friendship relations, which represent the simplest way to capture the concept of friendship on Facebook. On the other hand, there could be more refined representations of the Facebook social graph, such as taking into consideration the frequency of interaction among individuals of the network, to weight the importance of each tie. To this purpose, the feasibility of this study is complicated by the privacy issues deriving from accessing private information about users habits (such as the frequency of interaction with their friends), which limit our range of study.

Depending on this aspect, the second shortcoming of this study rises. In detail, the fact that we were concerned with the analysis of publicly accessible profiles implies that our sample only reproduces a partial picture of the Facebook social network which could slightly vary with respect to the overall social graph. To this purpose, another aspect which deserves more investigation is understanding how the incompleteness of the sampling affects the characteristics of the community structure.

**Conclusions**

The aim of this work was to investigate the emergence of social dynamics, organization patterns and mesoscopic features in the community structure of a large online social network such as Facebook. This task was quite thrilling and not trivial, since a number of theoretical and computational challenges raised.

First of all, we collected real-world data directly from the online network. In fact, as recently put into evidence in literature [40], the differences between synthetic and real-world data have profound implications on results.

After we reconstructed a sample of the structure of the social graph of Facebook, we unveiled its community structure. The main findings that emerged from the mesoscopic analysis of the community structure of this network can be summarized as follows:

(i) We assessed the tendency of online social network users to constitute communities of small size, proving the presence of a decreasing number of communities of larger size. This behavior explains the tendency of users to self-organization even in absence of a coordinated effort.

(ii) We investigated the occurrence of connections among communities, finding that some kind of links, commonly referred as to *weak ties*, are more relevant than others because they connect communities each other, according to the Granovetter's *strength of weak ties theory* [24] and in agreement with recent studies on other online social networks such as Twitter [21].

(iii) The community structure is highly clusterized and the diameter of the community structure meta-network is small (approximately around 4 and 5). These aspects indicate the presence of the *small world phenomenon*, which characterizes real-world social networks, according to sociological studies envisioned by Milgram [23] and in agreement with some heuristic evaluations recently provided by Facebook [18, 19].

The achieved results open space for further studies in different directions. As far as it concerns our long-term future research directions, we plan to investigate, amongst others, the following issues:

(i) Devising a model to identify the most representative users inside each given community. This would leave space for further interesting applications, such as the maximization of advertising on online social networks, the analysis of communication dynamics, spread of influence and information and so on.

(ii) Exploiting geographical data regarding the physical location of users of Facebook, to study the effect of strong and weak ties in the society [24]. In fact, is it known that a relevant additional source of information is represented by the geographical distribution of individuals [68–70]. For example, we suppose that strong ties could reflect relations characterized by physical closeness, while weak ties could be more appropriate to represent connections among physically distant individuals.

(iii) Concluding, we devised a strategy to estimate the strength of ties between social network users [71] and we want to study its application to online social networks on a large scale. In the case of social ties, this is equivalent to estimate the friendship degree between a pair of users by considering their interactions and their attitude to exchange information.

## Appendix

In this appendix we shortly discuss the background in community detection algorithms and explain the functioning of the two community detection methods adopted during our experimentation, namely LPA and FNCA.

### Community detection in complex networks

The problem of discovering the community structure of a network has been approached in several different ways. A common formulation of this problem is to find a partitioning $V = (V_1 \cup V_2 \cup \cdots \cup V_n)$ of disjoint subsets of vertices of the graph $G = (V, E)$ representing the network (in which the vertices represent the users of the network and the edges represent their social ties) in a meaningful manner.

The most popular quantitative measure to prove the existence of an emergent community structure in a network, called *network modularity*, has been proposed by Girvan and Newman [50, 51]. It is defined as the sum of the difference between the fraction of edges falling in each given community and the expected fraction if they were randomly distributed. Let consider a network which has been partitioned into *m* communities; its value of network modularity is

$$Q = \sum_{s=1}^{m} \left[ \frac{l_s}{|E|} - \left( \frac{d_s}{2|E|} \right)^2 \right] \qquad (6)$$

assuming $l_s$ the number of edges between vertices belonging to the *s*th community and $d_s$ the sum of the degrees of the vertices in the *s*th community. High values of *Q* imply high values of $l_s$ for each discovered community. In that case, detected communities are dense within their structure and weakly coupled among each other.

Partitioning a network in disjoint subsets may arise some difficulties. In fact, each user in the network possibly belongs to several different communities; the problem of overlapping community detection has recently received a lot of attention (see [34]). Moreover, may exist networks in which a certain individual may not belong to any group, remaining isolated, as recently put into evidence by Hunter *et al.* [16]. Such a case commonly happens in real and online social networks, as reported by recent social studies [72].

*Community detection techniques*
In its general formulation, the problem of finding communities in a network is solvable assigning each vertex of the network to a cluster, in a meaningful way. There exist different paradigms to solve this problem, such as the spectral clustering [73, 74] which relies on optimizing the process of cutting the graph, and the *network modularity* maximization methods.

Regarding spectral clustering techniques, they have an important limitation. They require a prior knowledge on the network, to define the number of communities present in the network and their size. This makes them unsuitable if the aim is to unveil the unknown community structure of a given network.

As for network modularity maximization techniques, the task of maximizing the objective function *Q* has been proved NP-hard [75], thus several heuristic techniques have been presented during the last years. The Girvan-Newman algorithm [50, 51, 76] is an example. It exploits the assumption that it is possible to maximize the value of *Q* deleting edges with a high value of betweenness, starting from the intuition that they connect vertices belonging to different communities. Unfortunately, the cost of this algorithm is $O(n^3)$, being *n* the number of vertices in the network; it is unsuitable for large-scale networks. A tremendous amount of improved versions of this approach have been provided in the last years and are extensively discussed in [30, 31].

From a computational perspective, some of the state-of-the-art algorithms are *Louvain method* [77, 78], LPA [48, 79], FNCA [49] and a voltage-based divisive method [80]. All these algorithms provide with near linear computational costs.

Recently, the problem of discovering the community structure in a network including the possibility of finding overlapping nodes belonging to different communities at the same time, has acquired a lot of attention by the scientists because of the seminal paper presented by Palla *et al.* [81]. A lot of efforts have been spent in order to advance novel possible strategies. For example, an interesting approach has been proposed by Gregory [82], that is based on an extension of the Label Propagation Algorithm adopted in this work. On the other hand, an approach in which the hierarchical clustering is instrumental to find the overlapping community structure has been proposed by Lancichinetti *et al.* [56, 83].

**Label Propagation Algorithm (LPA)**

The LPA (Label Propagation Algorithm) [48] is a near linear time algorithm for community detection. Its functioning is very simple, considered its computational efficiency. LPA uses only the network structure as its guide, is optimized for large-scale networks, does not follow any pre-defined objective function and does not require any prior information about the communities. Labels represent unique identifiers, assigned to each vertex of the network.

Its functioning is reported as described in [48]:

Step 1  To initialize, each vertex is given a unique label;

Step 2  Repeatedly, each vertex updates its label with the one used by the greatest number of neighbors. If more than one label is used by the same maximum number of neighbors, one is chosen randomly. After several iterations, the same label tends to become associated with all the members of a community;

Step 3  Vertices labeled alike are added to one community.

Authors themselves proved that this process, under specific conditions, could not converge. In order to avoid deadlocks and to guarantee an efficient network clustering, we accept their suggestion to adopt an *asynchronous* update of the labels, considering the values of some neighbors at the previous iteration and some at the current one. This precaution ensures the convergence of the process, usually in few steps. Raghavan *et al.* [48] ensure that five iterations are sufficient to correctly classify 95% of vertices of the network. After some experimentation, we found that this forecast is too optimistic, thus we elevated the maximum number of iterations to 50, finding a good compromise between quality of results and amount of time required for computation.

A characteristic of this approach is that it produces groups that are not necessarily contiguous, thus it could exist a path connecting a pair of vertices in a group passing through vertices belonging to different groups. Although in our case this condition would be acceptable, we adopted the suggestion of the authors to devise a final step to split the groups into one or more contiguous communities.

The authors proved its near linear computational cost [48].

**Fast Network Community Algorithm (FNCA)**

FNCA (Fast Network Community Algorithm) [49] is a modularity maximization algorithm for community detection, optimized for large-scale social networks.

Given an unweighted and undirected network $G = (V, E)$, suppose the vertices are divided into communities such that vertex $i$ belongs to community $r(i)$ denoted by $c_r(i)$; the function $Q$ is defined as Equation (7), where $A = (A_{ij})_{n \times n}$ is the adjacency matrix of network $G$. $A_{ij} = 1$ if node $i$ and node $j$ connect each other, $A_{ij} = 0$ otherwise. The $\delta$ function $\delta(u, v)$ is equal to 1 if $u = v$ and 0 otherwise. The degree $k_i$ of any vertex $i$ is defined to be

$k_i = \sum_j A_{ij}$ and $m = \frac{1}{2} \sum_{ij} A_{ij}$ is the number of edges in the network

$$Q = \frac{1}{2m} \sum_{ij} \left( \left( A_{ij} - \frac{k_i k_j}{2m} \right) \times \delta(r(i), r(j)) \right). \qquad (7)$$

We convert Equation (7) to Equation (8), which takes the function $Q$ as the sum of functions $f$ of all nodes. The function $f$ can be regarded as the difference between the number of edges that fall within communities and the expected number of edges that fall within communities, from the local angle of any node in the network. The function $f$ of each node can measure whether a network division indicates a strong community structure from its local point of view

$$Q = \frac{1}{2m} \sum_i f_i, \quad f_i = \sum_{j \in c_{r(i)}} \left( A_{ij} - \frac{k_i k_j}{2m} \right). \qquad (8)$$

The authors [49] proved that: (i) any node in a network can evaluate its function $f$ only by using local information (the information of its community); (ii) if the variety of some nodes label results in the increase of its function $f$ and the labels of the other nodes do not change, the function $Q$ of the whole network will increase too. The community detection algorithm used is based on these assumptions. It makes each node maximize its own function $f$ by using local information in the sight of local view, which will then achieve the goal that optimize the function $Q$.

Moreover, in complex networks with a community structure, holds true the intuition that any node should have the same label with one of its neighbors or it is itself a cluster. Therefore, each node does not need to compute its function $f$ for all the labels at each iteration, but just for the labels of its neighbors. This improvement not only decreases the time complexity of the algorithm, but also makes it able to optimize the function $Q$ by using only local information of the network community structure.

It has been proved that this algorithm, under certain conditions, could not quickly converge, thus we introduced an iteration number limitation $T$ as additional termination condition. Experimental results show that, the clustering solution of FNCA is good enough before 50 iterations for most large-scale networks. Therefore, iteration number limitation $T$ is set at 50 in all the experiments in this paper. Authors proved the near linear cost of this algorithm [49].

**Competing interests**
The author declares that he has no competing interests.

**Author's contributions**
EF designed and performed research, prepared figures, carried out empirical analysis, wrote and reviewed the manuscript.

**Acknowledgements**
The author is grateful to A. Flammini, B. Gonçalves, A. Lancichinetti, F. Menczer, F. Radicchi, and J.J. Ramasco for comments and suggestions on the manuscript.

**Endnotes**
<sup>a</sup> Data are represented in a compact format in order to save I/O operations and then are anonymized, in order not to store any kind of private data (such as the user-IDs).
<sup>b</sup> http://www.emilio.ferrara.name/datasets/.
<sup>c</sup> Remind that the vectors taken into account represent the communities of the network.
<sup>d</sup> To this regard, we put into evidence that the *x-axis* is reversed and we recall that the diameter of the considered community structures is 4.45 and 4.85, respectively for LPA and FNCA.

e  We recall that, being the network model adopted undirected, the meaning of source and target node is only instrumental to identify the end-vertex of each given edge.

f  The roles of weak ties is to connect small communities of acquaintances which are not that close to belong to the same community but, on the other hand, are somehow proficiently in contact.

g  https://sites.google.com/site/andrealancichinetti/cvis.